\documentclass[12pt]{article}
\usepackage{epsfig}
\begin{document}
\begin{titlepage}

\begin{center}
\hfill{FTUV-04-0407}

\hfill{IFIC-04-27}
\vspace{2cm}

{\Large  \bf Instanton induced quark dynamics and the pentaquark} \\
\vspace{0.50cm}
\renewcommand{\thefootnote}{\fnsymbol{footnote}}
N.I.Kochelev$^{a,b}$\footnote{kochelev@thsun1.jinr.ru},
H.-J. Lee $^{a}$\footnote{Heejung.Lee@uv.es},  V. Vento$^a$
\footnote{Vicente.Vento@uv.es}

\vspace{0.50cm}

{(a) \it Departament de F\'{\i}sica Te\`orica and Institut de F\'{\i}sica Corpuscular,\\
Universitat de Val\`encia-CSIC, E-46100 Burjassot (Valencia), Spain} \\

\vskip 1ex

{(b) \it Bogoliubov Laboratory of Theoretical Physics,\\
Joint Institute for Nuclear Research, Dubna, Moscow region, 141980
Russia}

\end{center}

\vskip 0.5cm

\centerline{\bf Abstract}

We analyze the existence of the exotic $\Theta^+$ from the
perspective of instanton induced quark dynamics. The 't Hooft
interaction gives strong attraction in specific channels of the
triquark $ud\bar s$ and diquark $ud$ configurations. In particular
it leads to a light $ud\bar s$ triquark cluster, with the mass
around $750\ {\rm MeV}$, in the $I=0$, $S=1/2$ and color 3 configuration,
and a light $ud$-diquark configuration, with mass $440\ {\rm MeV}$, in
the $I=0$, $S=0$ and color {$\bar{3}$} configuration. If we
consider the pentaquark as a bound state of such triquark and
diquark configurations in a relative $L=1$ state we obtain good
agreement with the data. The small width of $\Theta^+$ has a
natural explanation in this model.\vskip 0.5cm

\vskip 0.3cm \leftline{Pacs: 12.38.Aw, 12.38.Lg, 12.39.Ba, 12.39.-x}
\leftline{Keywords: quarks, instanton, hadrons, pentaquark}
\vspace{1cm}
\end{titlepage}

\section{Introduction}
The discovery of the exotic $\Theta^+$ baryon
\cite{nakano,barmin,stepanyan,newdat}, followed by recent
evidences of narrow pentaquark states with strangeness $S=-2$ and
charm \cite{NA49,HERA} has opened up a new scenario to understand
quark dynamics, in particular at low energies where non
perturbative mechanisms are expected.

There is a long history of predictions. The existence of exotics,
with the quark content $udud\bar{s}$, have been proposed in the
context of quark and bag models \cite{jaffe,strottman}. However
these states had large masses and typical hadronic widths. The
soliton model of baryons, based on the implementation of
spontaneous chiral symmetry breaking, was used for a successful
prediction of a very narrow exotic pentaquark state, the
$\Theta^+$, with the correct mass
\cite{praszalowicz,chemtob,diakonov1}, despite the fact that
recently some objections have been put forward to this analysis
\cite{rotation}.

After the detection of the pentaquark a plethora of calculations
have appeared aiming to understand the implications of its
existence in low energy quark dynamics. Models to describe the
complicated five particle scenario have been proposed
\cite{JW,KL,SZ,models,JM}. In particular some models, like the ones
proposed by Jaffe and Wilzek (JW), Karliner and Lipkin (KL), and
Shuryak and Zahed (SZ), consider that colored quark clusters
inside the pentaquark are formed and this explains its small mass
and width. Their approach leads to appealing simplified dynamical
schemes, but the need for quark clustering requires justification.
The aim of this letter is to prove that the quark dynamics
derivable from the instanton induced interaction justifies a
certain type of clustering. Also various lattice QCD calculations have
been carried out with contradictory results~\cite{lattice}.

The instantons, strong fluctuations  of gluon fields in the
vacuum, play a crucial  role in the realization of spontaneous
chiral symmetry breaking in Quantum Chromodynamics and in the
effective description of the spectroscopy for conventional
hadrons.
 The instantons induce the 't
Hooft interaction between the quarks which has strong flavor and
spin dependence, a behavior which  explains many features observed
in the hadron spectrum  and in hadronic
reactions (see reviews \cite{shurr,diakr,dorkochr} and references therein).

 Particularly relevant for us here is that, in the quark-quark
sector, the instanton induced interaction produces a strong
attraction in flavor antisymmetric states. The strength of this
interaction for the (ud) scalar diquark state is equal, for two
colors, to the the strength in the pion channel, the so-called
Pauli-G\"ursey symmetry, and only one-half weaker in the realistic
$N_c=3$ case \cite{SZ}. As a result of this dynamics a quasi-bound
very light $ud$--state can be formed. This mechanism implies that
models for the pentaquark with diquark correlations are preferable
to those  without any correlation between the quarks. Furthermore
the instanton induced interaction governs the dynamics between
quarks at intermediate distances, i.e. $r\approx \rho_c\approx 0.3
~{\rm fm}$, where $\rho_c$ is the average instanton size in the QCD
vacuum \cite{shuryak2}. This scale is much smaller than the
confinement size $R \approx 1~{\rm fm}$ and therefore it favors the
existence of clusters inside the large confinement region.

In this letter we consider a version of a triquark-diquark model
for the pentaquark motivated by the instanton induced interaction
between the quarks. We will show that taking into account the
strong instanton interaction in triquark and diquark clusters
allows us to understand the mass and width of the pentaquark.

\section{Perturbative and non-perturbative
interactions between quarks in multiquark hadrons}

Jaffe's  famous papers on multiquark states \cite{jaffem}, based
on the MIT bag model, motivated a wide discussion on the
properties of  exotic hadronic states. Most predictions have been
based on the assumption that the perturbative one-gluon exchange
interaction among quarks is the main mechanism to understand the
spectrum of multiquark systems. In an alternative approach
\cite{DKH,multiDK,oka}, the non-perturbative instanton induced
interaction has been suggested to  dominate the spin and flavor
dependent mass splitting between multiquark states and to provide
a very strong mixing between them.

The most important instanton induced interaction in quark systems
is the multiquark 't Hooft interaction, which arises from the
quark zero modes in the instanton field (see Fig.~1).
\begin{figure}[htb]
\centering \epsfig{file=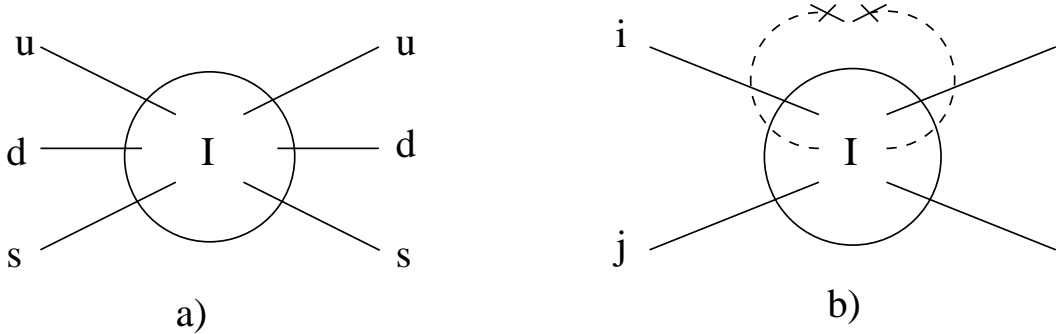,width=14cm} \vskip 1cm
\caption{ The  instanton induced a) three-quark $uds$ interaction
and b) two-quark $ud$, $us$, $ds$ interactions. In the figure $I$
denotes the instanton, $i,j=u,d,s, i\neq j$.}
\end{figure}

 For $N_f=3$~(Fig.~1a) and $N_c=3 $ this interaction is given by \cite{SVZ}:
\begin{eqnarray}
{\cal L}_{eff}^{(3)}&=&\int d\rho\
 n(\rho)\bigg\{\prod_{i=u,d,s}
\bigg(m_i^{cur} \rho-\frac{4\pi}{3}\rho^3\bar q_{iR}q_{iL}\bigg)
\nonumber\\
&&+\frac{3}{32}\bigg(\frac{4}{3}\pi^2\rho^3\bigg)^2
\bigg[\bigg(j_u^aj_d^a-\frac{3}{4}j_{u\mu\nu}^a
j_{d\mu\nu}^a\bigg)\bigg(m_s^{cur}\rho-\frac{4}{3}\pi^2\rho^3\bar
q_{SR}q_{sL}\bigg)\nonumber\\
&&+\frac{9}{40}\bigg(\frac{4}{3}\pi^2\rho^3\bigg)^2d^{abc}
j_{u\mu\nu}^aj_{d\mu\nu}^b
j_s^c+ {\rm perm.}\bigg]+\frac{9}{320}\bigg(\frac{4}{3}\pi^2\rho^3\bigg)^3
d^{abc}j_u^aj_d^b
j_s^c\nonumber\\
&&+\frac{if^{abc}}{256}\bigg(\frac{4}{3}\pi^2\rho^3\bigg)^3
j_{u\mu\nu}^aj_{d\nu\lambda}^b\relax
j_{s\lambda\mu}^c+(R\longleftrightarrow L) \bigg\} ,
\label{thooft3}
\end{eqnarray}
where, $m_i^{cur}$ is the quark current mass,
$q_{R,L}={(1\pm\gamma_5)q(x)/2}, \ j_i^a=\bar
q_{iR}\lambda^aq_{iL},{\ } j_{i\mu\nu}^a=\bar
q_{iR}\sigma_{\mu\nu}\lambda^aq_{iL}$,
 $\rho$ is the instanton size and  $n(\rho)$ is the density of
instantons \footnote{For quarks with nonzero virtualities $k_i^2$
the vertex (\ref{thooft3}) should be multiplied by factors $ Z_i=
F(k_i^2)$ for each incoming and outgoing quark legs. For small
values of the virtualities one can use the formula
$F(k_i^2)\approx e^{-\rho\sqrt{k_i^2}}$.}.

One can obtain an effective two-quark interaction induced by
instantons from the three-quark interaction (\ref{thooft3}) by
connecting two quark legs through the quark condensate (Fig.~1b).
In the limit of small instanton size one obtains simpler formulas
for effective two- and three-body point-like interactions
\cite{DKH,multiDK,oka}:
\begin{eqnarray}
{\cal H}_{eff}^{(2)}(r)&=&-V_2\sum_{i\neq j}\frac{1}
{m_im_j}\bar q_{iR}(r)q_{iL}(r) \bar
q_{jR}(r)q_{jL}(r)
\bigg[1+\frac{3}{32}(\lambda_u^a\lambda_d^a+{\rm perm.})\nonumber\\
&&+\frac{9}{32}(\vec{\sigma_u}\cdot\vec{\sigma_d}\lambda_u^a\lambda_d^a+{\rm perm.})\bigg]
+(R\longleftrightarrow L), \label{thooft2}
\end{eqnarray}
and
\begin{eqnarray}
{\cal H}_{eff}^{(3)}(r)&=&- V_3\prod_{i=u,d,s}\bar
q_{iR}(r)q_{iL}(r) \bigg[1+\frac{3}{32}(\lambda_u^a\lambda_d^a+{\rm perm.})
\nonumber\\
&&+\frac{9}{32}(\vec{\sigma_u}\cdot\vec{\sigma_d}\lambda_u^a\lambda_d^a+{\rm perm.})
 -\frac{9}{320}d^{abc}\lambda^a\lambda^b\lambda^c
(1-3(\vec{\sigma_u}\cdot\vec{\sigma_d}+{\rm perm.}))\nonumber\\
&&-\frac{9f^{abc}}{64}\lambda^a\lambda^b\lambda^c
(\vec{\sigma_u}\times\vec{\sigma_d})\cdot\vec{\sigma_s}\bigg]+(R\longleftrightarrow
L), \label{thooft4}
\end{eqnarray}
where $m_i=m_i^{cur} +m^*$ is the effective quark mass in the
instanton liquid. These forms are suitable for calculating the
instanton induced contributions within a constituent quark
picture.

In our estimates below, to avoid  uncertainties in the parameters
of instanton model (see recent discussion in \cite{SF}) and
uncertainties in the shape of quark wave functions, associated
with the confinement potential, we will treat the product of the
strength of four-quark instanton interaction $V_2$ and the
overlapping radial structures of the wave functions of the quarks
(\ref{thooft2}) as a free parameter, as suggested some time ago by
Shuryak and Rosner \cite{SR}. We will fix the value of this
parameter by fitting the masses of the hadronic ground states: the
baryon octet and decuplet, and the vector meson nonet. Thus, our
two body instanton interaction will have the form
\begin{equation}
V_{inst}^{qq}=-\sum_{i\neq j}\frac{a}
{m_im_j} \bigg[1+\frac{3}{32}(\lambda^a_u\lambda^a_d+{\rm perm.})
+\frac{9}{32}(\vec{\sigma_u}\cdot\vec{\sigma_d}\lambda_u^a\lambda_d^a
+{\rm perm.})\bigg],
\label{v2qq}
\end{equation}
for the quark-quark interaction. The color-spin structure of the
instanton induced quark-antiquark interaction can be obtained from
Eq.~(\ref{thooft2}) by crossing
\begin{equation}
V_{inst}^{q\bar q}=-\sum_{i\neq j}\frac{a} {m_im_j}
\bigg[1-\frac{3}{32}(\lambda_u^a\lambda^a_{\bar s}+{\rm perm.})
+\frac{9}{32}(\vec{\sigma_u}\cdot\vec{\sigma_{\bar s}}\lambda_u^a
\lambda_{\bar s}^a+{\rm perm.})\bigg], \label{v2qbarq}
\end{equation}
where
\begin{equation}
 \lambda_{\bar q}=-\lambda^*,\ \
\sigma_{\bar q}=-\sigma^* \label{barq}
\end{equation}
are color and spin generators for the antiquark representation.

In addition to the instanton interaction, we will also include in
the fit the perturbative one-gluon hyperfine interaction
\begin{eqnarray}
 V_{OGE}^{qq}&=-&\sum_{i>j}\frac{b}
{m_im_j}\vec{\sigma_i}\cdot\vec{\sigma_j}\lambda_i^a\lambda_j^a,
\label{qq2}
\end{eqnarray}
between quarks. For the quark-antiquark OGE interaction one should
use the substitution in Eq.~(\ref{barq}) \cite{jaffem}.

It is easy to show that three body instanton interaction does not
contribute to the masses of the ground state baryons. Therefore
its strength can not be fixed from such fit and should be
estimated within some model. The estimate of the three-body
instanton induced contribution to the mass of multiquark system
was considered for the first time in \cite{DKH},
 where its contribution to the mass of the H-dibaryon has been analyzed.
 This estimate was based on Shuryak's  version of the instanton
liquid model \cite{shuryak2}, in which the density of instantons
has  the form $n(\rho)\propto \delta(\rho-\rho_c)$. In this model
one can obtain a  relation between strengths of the three- and
two-body instanton induced interactions for zero quark
virtualities (see \cite{DKH}).
\begin{equation}
V_3=-V_2\frac{4\pi^2\rho_c^2}{3m_u m_d m_s} \label{ratio}
\end{equation}
We will use this relation below to estimate three-body
contribution to pentaquark mass.

\section{Masses of ground state hadrons}
We use the following mass formula for the ground hadronic states
\begin{equation}
M_h=E_0^{B,M}+\sum_i N_im_i+E_{I2}+E_{OGE}, \label{mass}
\end{equation}
where $N_i$ is number of the quarks with flavor $i$ in the state.
In Eq.~(\ref{mass})
\begin{eqnarray}
E_{I2}&=&<h|V_{I2}|h>=-\sum_{i\neq j}\frac{a}
{m_im_j}M_{i,j}^{I2}\nonumber\\
E_{OGE}&=&<h|V_{OGE}|h>=-\sum_{i>j}\frac{b} {m_im_j}M_{i,j}^{OGE}
\label{element}
\end{eqnarray}
 are the matrix elements of the
two-body instanton and OGE interactions. In comparison with the
Shuryak and Rosner constituent quark model with two-body instanton
induced interaction \cite{SR}, we have added the OGE contribution
and the parameter $E_0^{M,B}$. This new parameter represents the
contributions from the confinement forces and breaks the
additivity of the simple constituent quark model. For example, in
the MIT bag model approach this term would arise as a consequence
of the bag energy. We will assume that the value of this parameter
is the same for all hadrons with equal number of the valence
quarks. The values for the color-spin matrix elements
 for the hadronic ground states are shown in Table 1.
We  assume $m_u=m_d=m_0$ and  neglect the mixing between the
pseudoscalar octet and singlet meson states. In the vector meson
nonet the ideal mixing for their wave functions has been assumed.
\begin{table}
\begin{center}
\begin{tabular}{|c|c|c|c|c|c|} \hline
$Hadron$&$M_{00}^{OGE}$ &  $M_{0s}^{OGE}$  & $M_{ss}^{OGE}$
&$M_{00}^{I2}$ & $M_{0s}^{I2}$
\\ \hline\hline
$\pi$   & 16            &  0               &  0              & 6
& 0 \\ \hline $K$     &  0            & 16               &  0
& 0          & 6 \\ \hline $\eta$  & 16/3          &  0
&  32/3           & -2         & 8 \\ \hline
$\eta^{\prime}$& 32/3          &  0               &  16/3           & -4       & -8 \\
\hline $\rho/\omega $& -16/3          &  0               &  0
& 0       & 0 \\ \hline $K^*$   &  0            & -16/3
&  0              & 0          & 0 \\ \hline $\Phi$  &  0
&    0             & -16/3           & 0          & 0 \\ \hline $N
$  &  8            &  0               &  0              & 9/2
& 0 \\ \hline
$\Lambda $  &  16/3     &  8/3             &  0              & 3          & 3/2 \\
\hline $\Xi $     &  0        &  8               &  0
& 0          & 9/2 \\ \hline
$\Sigma $   &  0        &  8               &  0              & 0          & 9/2 \\
\hline $\Delta  $  & -8        &  0               &  0
& 0          & 0 \\ \hline
$\Sigma^* $ &  0        &  -8/3            &  -16/3              & 0          & 0 \\
\hline
$\Xi^* $    &  0        &  -8/3            &  -16/3              & 0          & 0 \\
\hline $\Omega $   &  0        &   0            &   -8
& 0          & 0 \\ \hline
\end{tabular}
\caption{Two-body color-spin matrix elements of one-gluon and
instanton interactions for ground state hadrons.} \label{matrix}
\end{center}
\end{table}
\begin{table}[htb]
\begin{center}
\begin{tabular}{|c|c|c|c|c|c|c|}\hline
$   $       &$\rho$ &  $\omega$  & $K^*$ &$\Phi$ & $N$ &$\Lambda$
\\ \hline\hline $M_h^0  $ & 740         &  740    &  884  & 1028
& 1218 &1362     \\ \hline $E_{OGE}$ & 19          &  19     &  12
& 8     & -29  &-25      \\ \hline $E_{I2}$&0           &  0
&  0    & 0     & -254 &-224      \\ \hline $M_h$  & 759
&  759    &  896  & 1036  & 935 & 1113     \\ \hline $M_h^{exp}$&
770        &  780    &  896  & 1020  & 940 &1116 \\ \hline $   $
&$\Sigma$ &  $\Xi$  & $\Delta$ &$\Sigma^*$ & $\Xi^*$ &$\Omega$  \\
\hline\hline $M_h^0 $ &1362     & 1506    & 1218     & 1362      &
1506    &1650  \\ \hline $E_{OGE}$&-19      & -19     & 29       &
22        & 16      &12    \\ \hline $E_{I2}$&-164   & -164
&  0       & 0          & 0 & 0\\ \hline $M_h$  & 1179      &
1323   &  1247    & 1384     & 1522 & 1662\\ \hline
$M_h^{exp}$&1192   &  1315   &  1236    & 1386     & 1532 & 1672
\\ \hline
\end{tabular}
\caption{The contributions to the baryon and vector meson nonet
masses $M_h$ that arise from the one-gluon exchange ( $E_{OGE}$),
the two-body instanton interaction ($E_{I2}$), the sum of quark
masses and the confinement energy contribution. $M_h^{exp}$ labels
the data.} \label{matrix1}
\end{center}
\end{table}

The result of our fit to the baryon and vector meson masses is
shown in Table 2. The values for the parameters are
\begin{eqnarray}
m_0&=&263\ {\rm MeV},{\ } m_s=407\ {\rm MeV}, {\ }E_0^M=214\ {\rm MeV},
\nonumber\\
E_0^B&=&429\ {\rm MeV},{\ } a=0.0039\ {\rm GeV}^3, {\  }b=0.00025\ {\rm GeV}^3.
\label{par}
\end{eqnarray}
From Table 2 one can conclude that the one-gluon exchange
interaction contributes little to the hadron masses \footnote{The
value of strength of one-gluon exchange in Eq.~(\ref{par})
corresponds to $\alpha_s\approx 0.4$, if one uses MIT bag model
quark wave functions with a bag radius $R\approx 1$fm.} and the
main contribution to the spin-spin splitting between hadron
multiplets comes from the instanton induced interaction. This
conclusion is in agreement with the results of the constituent
quark model calculations with instanton forces using various forms
of quark wave functions \cite{DKMIT,oka2,petry}. This
result was also confirmed independently by the calculation of
instanton effects on hadron masses within the QCD sum rule
approach \cite{ISUM}. With the values of the parameters shown in
(\ref{par}), the masses of the pseudoscalar mesons are the
following
\begin{eqnarray}
m_{\pi}&=&344\ {\rm MeV} \;(140\ {\rm MeV}), {\ } m_{K}=628\ {\rm MeV} \,
(498\ {\rm MeV}),\nonumber\\
m_{\eta}&=&709\ {\rm MeV} \;(550\ {\rm MeV}),{\ } m_{\eta^\prime}=1302\
{\rm MeV} \;(960\ {\rm MeV}), \label{pseudo}
\end{eqnarray}
where in parenthesis we wrote their experimental values.

One can see, that our model overestimates the masses of particles
from the pseudoscalar nonet. This happens because we are neglecting the
difference in size between the pseudoscalar and vector nonet
mesons in our model. It is known that the one-gluon exchange
contribution behaves as $\approx 1/R$ \cite{MIT}, while the
instanton contribution as $\approx 1/R^3$ \cite{DKMIT}. In the
case of pseudoscalar octet both one-gluon and instanton exchanges
give very strong attraction between the constituents, due to the
large value of the color-spin matrix elements (see Table 1). This
attraction leads to a small effective size for the quark-antiquark
system. In fact, the size of the such systems should be comparable
to the instanton size. The $\eta^\prime$, on the other hand,
results from a different behavior, the instantons give very strong
repulsion in this channel and therefore the size of the
$\eta^\prime$ must be larger than that of the  vector
meson\footnote{A detailed discussion of the $\eta^\prime$
properties are beyond the scope of this article. We just mention
that the instanton contribution leads to a large mass splitting
between the $\eta$ and $\eta^\prime$. Therefore, the $U_A(1)$
problem, does not arise in the instanton model.}.

We parametrize the size dependence of the particles by a new
parameter $r=R_{eff}/R$ which affects the instanton and one-gluon
interactions. We take $R\approx 1$fm, the size of the
conventional baryons. Thus, in (\ref{element}) for the
pseudoscalar mesons we multiply the one-gluon exchange
contribution by $1/r$  and the instanton contribution by $1/r^3$.
Moreover for those systems with a size comparable with the
instanton size, one should introduce the instanton form factor,
\begin{equation}
F(r)\approx e^{-2N\rho_c/R_{eff}}, \label{form2}
\end{equation}
where $\approx 1/R_{eff}$ is the average quark virtuality in the
system and  $N=2$ or $N=3$ for the instanton induced two- and three-body
 interactions, respectively.
The result of the new fit to the masses of the pseudoscalar octet
gives
\begin{eqnarray}
m_{\pi}&=&112\ {\rm MeV}, {\ } m_{K}=498\ {\rm MeV}, {\ }
m_{\eta}=581\ {\rm MeV}
\label{pseudo2}
\end{eqnarray}
in a good agreement with the data. As a result of the fit we
observe that the effective size, of systems with strong instanton
attraction, becomes
\begin{equation}
R_{eff}\approx 0.72 R,
\end{equation}
where conventionally $R \approx 1$fm. Of course, this estimate is
rather rough, but it shows that one can expect that the quark
systems with strong instanton attraction are small compared to the
others \footnote{In the bag model the size of the system is
determined by the position of minimum of the hadronic mass as a
function of bag radius $R$ and a strong instanton interaction
shifts downward the position of the minimum.}.

\section{Diquark-triquark model for pentaquark and instantons}

In the previous section the strong influence of the instantons on
the dynamics of the colorless quark systems has been shown. Here
the application of  the model to the color quark systems with
instanton attraction within the five quark pentaquark system  will
be considered.

Let us start with the discussion of the $\Theta^+$ $udud\bar s$
wave function in the model with instanton induced correlations
between quarks. The observed $\Theta^+$ state is very light in
comparison with the expectation of the constituent quark model for
the typical mass of the $udud\bar s$ system \footnote{One can
estimate it by using our fit above $M(udud\bar s)\approx
E_0^B+E_0^M+4m_0+m_s\approx 2100~{\rm MeV}$, which is much larger then
experimental value $M_{\Theta^+}\approx 1540~{\rm MeV}$.}, and has a
very small width. Thus, we should expect a non trivial wave
function for the pentaquark. One of the peculiarities of the
instanton induced interaction is its strong flavor dependence,
i.e., it is not vanishing  only for the interaction among quarks
of  different flavor. For the $ud$ diquark system the strong
instanton attraction is possible only in isospin $I=0$ channel.
Thus, preferably the configuration in the $udud$ subsystem will be
two separated isoscalar $ud$ diquarks. The remaining antiquark
$\bar s$ can join one of the diquarks to create a triquark $ud\bar
s$ configuration in the instanton field. In this triquark state
all quarks have different flavors, therefore the instanton
interaction is expected to be  maximal. Another peculiarity of the
instanton interaction is that it is maximal in the system with the
minimal spin. Therefore, a pentaquark configuration with $S=1/2$
$ud\bar s$ triquark and $ud$ $S=0$ diquark should be preferable.
Therefore our final triquark-diquark picture for the pentaquark
with instanton forces between quarks arises as shown in Fig.2a,
where the triquark is a quasi-bound state in the field of the
instanton (antiinstanton) and the diquark is a quasi-bound state
in the antiinstanton (instanton) field \footnote{There is
attraction (repulsion) between pseudoparticles with the same
(opposite) topological charge. Therefore the
instanton-antiinstanton (IA) configuration has smaller energy then
the II and AA configurations.}. To avoid the coalescence of the
triquark-diquark state into the single cluster $udud\bar s$
configuration, where the instanton interaction is expected to be
much weaker, due to the Pauli principle for the same flavor quarks
in instanton field, we assume a non-zero orbital momentum $L=1$ in
the triquark-diquark system. The centrifugal barrier protects the
clusters from getting close and prohibits the formation of the
much less bound five quark cluster.

It should be mentioned, that, from our point of view, the
possibility of a pentaquark configuration formed by two
$ud$-diquark clusters and a single antiquark $\bar s$, shown in
(Fig.~2b), as implied by the JW and SZ models, is suppressed by
extra powers of the instanton density,
$f=n_{eff}\pi^2\rho_c^4\approx 1/10$ in the instanton model as
compared with the triquark--diquark configuration of Fig.~2a.
\begin{figure}[htb]
\centering \epsfig{file=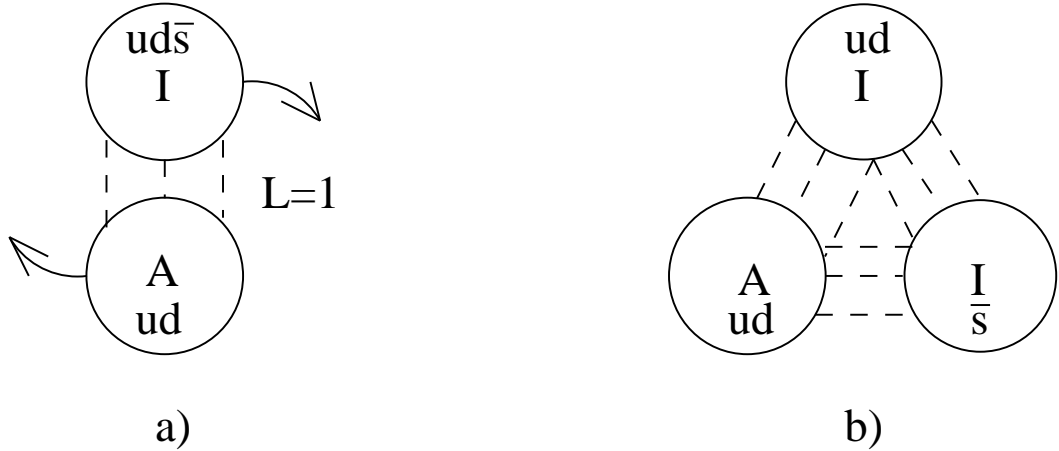,width=14cm} \vskip 1cm
\caption{ (a) Our instanton model for the pentaquark, (b) is the
instanton picture for JW and SZ models. $I$ ($A$) denote instanton
 (antiinstanton) configurations. Dashed lines indicate gluon lines.}
\end{figure}

According to Pauli statistics  in the $ud\bar s$ $I=0$ triquark
 state the $ud$ diquark can be
in $S=0$ spin and  $\bar 3_c$ color state (A state) or in $S=1$,
$6_c$ (B state). In KL only B has been considered. In fact, there
is a strong mixing between the two states due to both the
one-gluon and instanton interactions (see below), and one cannot
neglect either.
\footnote{Other types of mixing have also been discussed in the
literature~\cite{JM}.}
In Table 3, the diagonal ($\langle {\rm A}|H|{\rm A}\rangle,
\langle {\rm B}|H|{\rm B}\rangle$) and
the non-diagonal ($\langle {\rm A}|H|{\rm B}\rangle$) color-spin matrix
elements  of the one-gluon and instanton interactions for the $ud\bar s$
triquark and scalar-isoscalar $ud$ diquark are shown. For the instanton
interaction we also have included the three-body color-spin
$M^{I3}$ matrix elements of the interaction (\ref{thooft4}).
Its explicit form  for the $ud\bar s$ state is
\begin{eqnarray}
E^{I3}_{ud\bar s}&=& b_3
\bigg[\ 1+\frac{3}{32}(\lambda_u^a\lambda_d^a-\lambda_u^a\lambda_{\bar s}^a
-\lambda_d^a\lambda_{\bar s}^a )+
\frac{9}{32}(\vec{\sigma_u}\cdot\vec{\sigma_d}\lambda_u^a\lambda_d^a+{\rm perm.})
\nonumber\\
&&\hspace{0.6cm}+\frac{9}{320}d^{abc}\lambda_u^a\lambda_d^b\lambda_{\bar s}^c
(1-3(\vec{\sigma_u}\cdot\vec{\sigma_d}-\vec{\sigma_u}\cdot\vec{\sigma_{\bar
s}} -\vec{\sigma_d}\cdot\vec{\sigma_{\bar s}}))
\nonumber\\
&&\hspace{0.6cm}-\frac{9}{64}f^{abc}\lambda_u^a\lambda_d^b\lambda_{\bar s}^c
(\vec{\sigma_u}\times\vec{\sigma_d})\cdot\vec{\sigma_{\bar s}}\ \bigg],
\label{I3}
\end{eqnarray}

\begin{table}[htb]
\begin{center}
\begin{tabular}{|c|c|c|c|c|c|} \hline
$State$&$M_{00}^{OGE}$ & $M_{0s}^{OGE}$ & $M_{00}^{I2}$ &$M_{0s}^{I2}$ & $M^{I3}$ \\
\hline\hline $ud$   & 8             &  0             &  3
& 0          & 0 \\ \hline $\langle{\rm A}|H|{\rm A}\rangle$&8            &  0
&  3           & 9/4        & 9/2 \\ \hline $\langle{\rm B}|H|{\rm B}\rangle$& 4/3
&  40/3          &  3/2         & 51/8       & 27/4 \\ \hline
$\langle{\rm A}|H|{\rm B}\rangle$& 0           &  $-(96)^{1/2}$    &  0
&$-(243/32)^{1/2}$ & $-(243/8)^{1/2}$ \\ \hline
\end{tabular}
\caption{Color-spin matrix elements of the one-gluon and instanton
interactions for triquark $ud\bar s$ and diquark $ud$ states}
\label{matrix2}
\end{center}
\end{table}
From Table 3 it can be seen that the one-gluon and instanton two
body interactions give strong attraction in the diquark and
triquark channels. For example, comparison of the matrix elements
in Table 1 and Table 3, in the $SU(3)_f$ symmetry limit, show
that, in the B triquark state, the attraction is even larger than
in the case of the Goldstone pion! Therefore, one can expect very
light triquark cluster type configurations whose size is that of
the pion. We see in Table 3 that the off-diagonal matrix elements
between the states A and B are rather large and the physical
states arise as a mixing of these states. Three-body instanton
induced forces give a repulsion in both A and B states due to the
opposite sign of their strength, (\ref{ratio}). The final
result depends on the overlap between the $ud\bar s$ quark wave
functions. We estimated three-body contribution by using the bag
model wave functions. The final result is
\begin{equation}
\Delta M_{I3}\approx -0.03M^{I3}\Delta M_{I2}^N
\frac{4\pi^2\rho_c^2}{3m_sR^3r^6}e^{4\rho_c/R(1-3/2r)},
\label{tree}
\end{equation}
where $\Delta M_{I2}^N$ is the instanton contribution to the
nucleon mass, $r=R_{tri}/R$ and we take into account the
difference in form factors for the two- and three-body
interactions, Eq.~(\ref{form2}). We assume that due to the similarity
in strength of the instanton attraction in the pseudoscalar octet,
triquark and diquark configurations, the size of the all these
states is the same. Thus we may use the same value for parameter
$r$
\begin{equation}
r=R_{eff}/R\approx R_{tri}/r\approx R_{di}/R.
\end{equation}
Finally  we have
\begin{eqnarray}
& &\bullet\ {\rm diquark}: M_{di}=442\ {\rm MeV}, {\ } M_{0di}=740\ {\rm MeV},
\nonumber\\
& & \hspace{2cm}\Delta M_{OGE}=-24\ {\rm MeV},  {\ } \Delta M_{I2}=-274\ {\rm MeV}; \nonumber\\
& &\bullet\ {\rm triquark\ A} : M_{tri}=955\ {\rm Mev}, {\ } M_{0tri}=1362\ {\rm MeV},
\nonumber\\
& &\hspace{2cm}\Delta M_{OGE}=-40\ {\rm MeV},  \Delta M_{I2}=-407\ {\rm MeV}, {\ }
 \Delta M_{I3}= 40\ {\rm MeV};\nonumber\\
& &\bullet\ {\rm triquark\ B} : M_{tri}=859\ {\rm MeV}, {\ } M_{0tri}= 1362\ {\rm MeV},
\nonumber\\
& &\hspace{2cm}\Delta M_{OGE}=-50\ {\rm MeV}, {\ } \Delta M_{I2}= -513\ {\rm MeV},
{\ } \Delta M_{I3}= 60\ {\rm MeV};\nonumber\\
& &\bullet\ {\rm off-diagonal\ AB} : \Delta M_{OGE}= 32\ {\rm MeV}, {\ } \Delta M_{I2}=
164\ {\rm MeV}, {\ }\nonumber\\
& &\hspace{2cm} \Delta M_{I3}=-49\ {\rm MeV}, \label{m2}
\end{eqnarray}
where $M_0$ is the mass of the state without the one-gluon and
instanton contributions. From (\ref{m2}) it follows that the
two-body instanton instanton interaction  gives a very large and
negative contribution to the masses for all diquark and triquark
states. At the same time, the one-gluon contribution is rather
small. After diagonalization of the mass matrix for the A and B
states, we obtain for two mixed triquark states
\begin{equation}
M^{tri}_{light}=753\ {\rm MeV} {\ } {\rm and}
{\ } M^{tri}_{heavy}=1061\ {\rm MeV}.
\label{trimass}
\end{equation}
Comparing  the masses of non-mixed  (\ref{m2}) and mixed
(\ref{trimass}) states we see that the mixing is an important
effect in the spectroscopy of the triquark states. It increases
the difference between the two states from $96\ {\rm MeV}$ to $308\ {\rm MeV}$,
producing a very light $ud\bar s$ triquark state with a mass $753\ {\rm MeV}$.
It is about $360\ {\rm MeV}$ more bound than the lightest $uds$
$\Lambda$ state. The reason is simple. Both the one-gluon and
instanton interactions are twice more attractive in the
quark-antiquark channel than in the quark-quark case. The mass of
light triquark cluster is smaller then the sum of masses of the K
meson and the constituent u and d quarks. Therefore, the
pentaquark cannot dissociate to the Ku(d) system. Thus, the
$\Theta^+$, as a system of light triquark and diquark clusters,
can decay only by rearrangement of the quarks between these
clusters. However, this rearrangement is highly suppressed by the
orbital momentum $L=1$ barrier between the clusters.
 As a consequence, the centrifugal barrier, provides the mechanism
for a very small width in the case of the $\Theta^+$. The other,
heavy triquark , $1061\ {\rm MeV}$ state (\ref{trimass}), can easily
dissociate to the Ku(d) system. For this state, which should be
approximately $300\ {\rm MeV}$ above of $\Theta^+$ state, a very large
width is expected. We have found also a rather small mass $442\ {\rm MeV}$
for the $S=0$, $I=0$ $ud$-diquark. This mass is in agreement
with the estimate of $\approx 420\ {\rm MeV}$ for this diquark obtained
within the QCD sum rule approach using instanton induced
interactions \cite{SSV}.

Finally let us estimate the total mass of $\Theta^+$ if built as a
system of a triquark cluster with mass $753\ {\rm MeV}$, a diquark
cluster with mass $442 MeV$ bound together in relative  $L=1$
orbital momentum state. The reduced mass for such triquark-diquark
system is $M_{red}^{tri-di}=279\ {\rm MeV}$. This mass is approximately
equal to the "effective" reduced mass of the strange quarks in the
$\Phi$ meson, $M_{red}^{\Phi}\approx M_{\Phi}/4=255\ {\rm MeV}$. For two
strange quarks, the $L=1$ energy of orbital excitation, can be
estimated from the experimental mass shift between $\Phi$ meson
and the $L=1$ $f_1(1420)$ state
\begin{equation}
\Delta E(L=1)\approx M_{f_1(1420)}-M_{\Phi}=400\ {\rm MeV}. \label{L1}
\end{equation}
By neglecting the small difference between the reduced mass in
strange-antistrange quark system and triquark-diquark system, we
estimate the mass of the light pentaquark in our triquark-diquark
cluster model with instanton interaction as
\begin{equation}
M_{\Theta^+}=M_{light}^{tri}+M_{di}+\Delta E(L=1)\approx 1595\ {\rm MeV},
\label{theta}
\end{equation}
which is close to the data.

We should mention that our estimate of the $L=1$ excitation energy
is larger by a factor of two than the KL estimate ($207\ {\rm MeV}$)
\cite{KL}. The KL estimate has been obtained from the assumption
that due to the approximately equal reduced mass of the
triquark-antidiquark and $c\bar s$ systems, this energy is equal
to the $L=1$ excitation energy in $D_s$  mesons. An additional
assumption was done in interpreting the new $D_s(2317)$ state
\cite{expD} as a $0^+$ excitation of $0^-D_s(1969)$. We don't want
to discuss here the rather controversial status of the $D_s(2317)$
in the constituent quark model \footnote{It is rather difficult to
explain the small mass of this meson in the constituent quark
model (see references in \cite{expD}.)}, we would like to emphasize
simply that the reduced mass for the triquark-diquark system of
$458\ {\rm MeV}$ in their equation (2.4) was been obtained without the
contribution from the hyperfine interaction. After including this
effect one gets for the triquark state $890\ {\rm MeV}$ and  $495\ {\rm MeV}$
for mass of the diquark state. As a result, the corrected reduced
mass triquark-diquark system in KL model is $318\ {\rm MeV}$ which is
much smaller then reduced mass of $c\bar s$ system $410\ {\rm MeV}$ in
the KL paper. We can estimate the orbital excitation energy for
the KL model by using simple dependence of this energy on the
reduced mass of the system \cite{ROSNER}\footnote{We are grateful Sergo
Gerasimov for discussion the problems of estimating the energy of
orbital excitations in the constituent quark model.}
\begin{equation}
\Delta E(L=1)\propto M_{red}^{-n/(n+2)}, \label{ex}
\end{equation}
which one obtains from the solution of the Schr\"odinger equation
in a confinement potential $\propto r^n$. By putting conveniently
$n=1$ in (\ref{ex}) and using experimental information on the
$L=1$ excitation energy in the $s\bar s$ system, we estimate
\begin{equation}
\Delta E(L=1)_{KL}^{corrected}\approx 370\ {\rm MeV}. \label{ex2}
\end{equation}
This value is much larger then original KL estimate of $207\ {\rm MeV}$
and should lead to a significant deviation of their final result
for the mass of $\Theta^+$ from the experiment.

\section{Conclusion}

We have suggested a triquark-diquark model for the pentaquark
based on instanton induced interaction. It is shown that  this
strong interaction, which is at the origin  of the light
pseudoscalar octet of the mesons, leads also to the very light
$ud\bar s$ triquark and $ud$ diquark color states. As the result,
the possibility to explain the smallness of both, mass and width,
for the observed $\Theta^+$ based on triquark-diquark model with
strong instanton attraction between quarks, has been shown.

Let us discuss another possible signals for existence of very
light $ud\bar s$ triquark state. One interesting multiquark system
with expected small width can be a triquark-antitriquark system
with non-zero orbital momentum. The estimates of the mass of such
system in $L=1$ state within our model gives the number
\begin{equation}
M_{tri-antitri}=2M_{light}^{tri}+\Delta E(L=1, M_{red}=377\ {\rm MeV})
\approx 1860\ {\rm MeV} \label{baryonium}
\end{equation}
This mass is slightly smaller that two nucleon masses $2M_N=1880\ {\rm MeV}$
and therefore this triquark-antitriquark state  can  provide
a new explanation of the large near threshold enhancement in
$p\bar p$ spectrum  in the reaction $J/\Psi\rightarrow \gamma p\bar
p$ found by the BES Collaboration \cite{BES}.

One also can consider a system consisting of a light $(ud\bar
s)$-triquark and a flavor antisymmetric $us$ or $ds$ diquark.
According  of our model the  $us$ and $ds$ diquarks should be
heavier by $250\ {\rm MeV}$ and therefore its mass should be around
$1800\ {\rm MeV}$. We should mention, that in our model we do not expect very
narrow multiquark states with $u\bar d s$ or $d \bar u s$-triquark
state clusters inside. The reason is simple. Due to small mass of
the pion this triquark can easy dissociate to pion and a
constituent strange quark \footnote{ The first evidence for the
candidates for the narrow  pentaquark states with quark content
$d\bar usds$ and $u\bar dsds$ was found by NA49 Collaboration
\cite{NA49}. These results have been criticized in
\cite{fischer}.}.

We should emphasize, that  due to the specific properties of the
$ud\bar s$ light quark state, it should play the important role
not only in the spectroscopy of the multiquark states, but also in
different hadronic reactions.
 This triquark may also give the rise to  properties of
the quark-gluon plasma and nuclear matter.

\begin{center}

{\bf Acknowledgments}

\end{center}

We are grateful to A.E.Dorokhov and S.B.Gerasimov
for useful comments. N. I. Kochelev is  grateful to the
University of Valencia for the warm hospitality and financial
support. V. Vento thanks CERN-TH for the hospitality during the
early stages of this work.  H.-J. Lee is a Postdoctoral fellow
from SEEU-SB2002-0009. This work was supported by grants
MCyT-FIS2004-05616-C02-01 and GV-GRUPOS03/094~(VV), and
RFBR-04-02-16445, RFBR-03-02-17291 and INTAS-00-00-366 (NIK).

\end{document}